\documentclass[twocolumn,showpacs,preprintnumbers,nofootinbib,amssymb]{revtex4-1}

\usepackage{epsfig, amssymb,graphicx} 
\usepackage{amsmath, amsfonts}
\usepackage{bm} 

\usepackage[linktocpage]{hyperref}
\usepackage[caption=false]{subfig}
\usepackage[usenames]{color}

\def\nn{\nonumber}

\def\be{\begin{equation}}
\def\ee{\end{equation}}
\def\beq{\begin{eqnarray}}
\def\eeq{\end{eqnarray}}

\def\r{\right}
\def\l{\left}

\begin{document}

\title{Light rings as observational evidence for event horizons:\\ Long-lived modes, ergoregions and nonlinear instabilities of ultracompact objects}

\author{Vitor Cardoso} 
\affiliation{CENTRA, Departamento de F\'{\i}sica, Instituto Superior T\'ecnico, Universidade de Lisboa,
Avenida Rovisco Pais 1, 1049 Lisboa, Portugal}
\affiliation{Perimeter Institute for Theoretical Physics Waterloo, Ontario N2J 2W9, Canada}
\affiliation{Department of Physics and Astronomy, The University of Mississippi, University, MS 38677, USA}

\author{Lu\'is C. B. Crispino}
\affiliation{Faculdade de F\'{\i}sica, Universidade Federal do Par\'a, 66075-110, Bel\'em, Par\'a, Brazil}

\author{Caio F. B. Macedo}
\affiliation{Faculdade de F\'{\i}sica, Universidade Federal do Par\'a, 66075-110, Bel\'em, Par\'a, Brazil}

\author{Hirotada Okawa} 
\affiliation{CENTRA, Departamento de F\'{\i}sica, Instituto Superior T\'ecnico, Universidade de Lisboa,
Avenida Rovisco Pais 1, 1049 Lisboa, Portugal}

\author{Paolo Pani} 
\affiliation{CENTRA, Departamento de F\'{\i}sica, Instituto Superior T\'ecnico, Universidade de Lisboa,
Avenida Rovisco Pais 1, 1049 Lisboa, Portugal}
\affiliation{Dipartimento di Fisica, ``Sapienza'' Universit\`a di Roma, P.A. Moro 5, 00185, Roma, Italy}


\begin{abstract} 
Ultracompact objects are self-gravitating systems with a light ring.
It was recently suggested that fluctuations in the background of these objects are extremely long-lived and might turn unstable at the nonlinear level, if the object is not endowed with a horizon.
If correct, this result has important consequences: objects with a light ring are black holes.
In other words, the nonlinear instability of ultracompact stars would provide a strong 
argument in favor of the ``black hole hypothesis,'' once electromagnetic or gravitational-wave observations confirm the existence of light rings.
Here we explore in some depth the mode structure of ultracompact stars, in particular constant-density stars and gravastars.
We show that the existence of very long-lived modes --~localized near a second, stable null geodesic~-- is a generic feature of gravitational perturbations of such configurations. Already at the linear level, such modes become unstable if the object rotates sufficiently fast to develop an ergoregion.
Finally, we conjecture that the long-lived modes become unstable under fragmentation via a Dyson-Chandrasekhar-Fermi mechanism at the nonlinear level.
Depending on the structure of the star, it is also possible that nonlinearities lead to the formation of small black holes close to the stable light ring. Our results suggest that the mere observation of a light ring is a strong evidence for the existence of black holes.
\end{abstract}

\maketitle

\section{Introduction}

Our current understanding of stars and stellar evolution strongly suggests that
sufficiently compact, massive objects are unstable against gravitational collapse.
Neutron stars, with compactness $2GM/c^2R\sim 1/3$ cannot sustain masses larger than $\sim 3M_{\odot}$, 
whereas giant stars with masses $M\gtrsim 10 M_{\odot}$ have compactnesses orders of magnitude smaller. In other words, ordinary matter cannot support the enormous self-gravity of a massive and ultracompact object, so that the latter is naturally expected to be a black hole (BH).

The above picture has been challenged by the construction of exotic objects relying on different support mechanisms. For example, boson stars made up of fundamental massive scalar fields can be as compact as a neutron star and as massive as the BH candidate at the center of our galaxy~\cite{Liebling:2012fv,Macedo:2013qea}. Several other --~albeit more artificial~-- objects such as gravastars~\cite{Mazur:2001fv}, superspinars, etc, share similar properties~\cite{Cardoso:2007az,Cardoso:2008kj} and have been proposed as prototypical  alternatives to stellar and massive BHs. 

The observation --~or lack thereof~-- of a surface would be bullet-proof indication that compact dark objects have star-like properties or are instead endowed with an event horizon. Such tests are extremely challenging to perform in the optical window, but will become available with the advent of gravitational-wave astronomy: the oscillation modes of BHs have a very precise and well-known structure, which can be tested against observations~\cite{Berti:2005ys,Berti:2009kk,Berti:2006qt}, while the presence of a surface should be imprinted also on the gravitational waves generated during the merger of two objects~\cite{Kesden:2004qx,Macedo:2013qea,Pani:2009ss} (but see the discussion in Sec.~\ref{QNMstraw}).

Fortunately, General Relativity also comes to the rescue in helping to discriminate the nature of compact objects. Very compact
and highly spinning objects with an ergoregion but without a horizon are unstable~\cite{1978CMaPh..63..243F}. Thus, rapidly-spinning compact objects 
must, in principle, be black holes~\cite{Cardoso:2007az,Cardoso:2008kj}. However, observations of these objects are marred with uncertainties and not all of them are highly spinning. Furthermore, depending on the compactness and the spin, the instability time scale might be longer than the age of massive objects~\cite{Chirenti:2008pf}, making it an ineffectual mechanism.

Very recently, a new mechanism was put forward that excludes {\it any} ultracompact star configuration on the grounds that such object would be nonlinearly unstable~\cite{Keir:2014oka}.
If correct, this mechanism would close the ``BH paradigm'' project: within General Relativity, the observation of an ultracompact object
would be an observation of a BH~\footnote{We are assuming that the instability time scale is short enough to dominate the dynamical evolution of the compact object, see below for a discussion.}. The relevance of such corollary calls for a detailed analysis of the decay of linear perturbations in the spacetime of ultracompact configurations, and of the nonlinear evolution of such objects.
Here, we wish to take a first step in this direction by studying linear perturbations.

We show that linear perturbations of \emph{any} ultracompact star do become arbitrarily long-lived in the eikonal regime,
and correspond to fluctuations trapped between the outer, unstable light ring and the origin. Such modes are peaked at the location of a \emph{stable} light ring, whose existence is a peculiar property of these ultracompact objects. Already at the linear level, these long-lived modes turn unstable against the ergoregion instability~\cite{1978CMaPh..63..243F} when a small amount of rotation is added to the star.
Furthermore, at the nonlinear level, we provide evidence that the outer layers of the star may fragment and subsequently fallback on the star's core, making it 
dynamically resemble a ``boiling object''. Consequent emission of gravitational radiation will cause mass loss and a decrease in compactness,
leading to stable stars without light rings. Depending on the star structure, fragmentation could even be due to BH formation, in which case the end-state is a BH.

\section{Ultracompact objects}
We define an ultracompact object as one possessing a light ring (in addition, we will be working mostly with horizonless objects). We focus here on static, spherically symmetric spacetimes described by (henceforth we use geometrical units $G=c=1$)
\begin{equation}
ds^2=-f(r)dt^2+B(r)dr^2+r^2d\Omega_2^2\,. \label{metricspherical}
\end{equation}
If we use coordinates where the spacetime is manifestly asymptotically flat, then $f(r),\,B(r)\to 1$ at large distances.
Moreover, the requirement that the spacetime be locally flat and regular implies that $f(r)$ and $B(r)$
be finite at the origin $r=0$ for any horizonless object.

The radial equation for null geodesics in this geometry reads~\cite{Cardoso:2008bp}
\be
B(r) f(r)\dot{r}^2=E^2-V_{\rm geo}\equiv E^2-L^2\frac{f(r)}{r^2}\,,
\ee
where $V_{\rm geo}$ is the geodesic potential\footnote{To simplify the comparison with the effective potential for wave propagation, here we defined the geodesic potential $V_{\rm geo}=E^2-B(r)f(r) V_r$, where $V_r$ is the effective potential adopted in Eq.~(29) of Ref.~\cite{Cardoso:2008bp}.} and $E$ and $L$ are the conserved specific energy and angular momentum of the geodesic.
The existence of one (unstable) light ring for ultracompact objects --~at roughly $r_{\rm LR}\sim 3M$ for spherically symmetric configurations~--
means that $V_{\rm geo}$ has a local maximum at that point. Because $V_{\rm geo}$ diverges and is positive at the origin for ultracompact stars, this also implies
the existence of a local minimum and therefore of {\it a second} --~stable~-- light ring, typically within the star. 

The existence of a stable light ring is thus an unavoidable feature of any ultracompact star and has dramatic consequences for the dynamics of the latter.
Indeed, a stable light ring suggests that some modes can become very long-lived~\cite{Comins,Cardoso:2008bp,Barausse:2014tra,Barausse:2014pra}.
When this happens, nonlinear effects can become important and destabilize the system. In a nutshell, this was the argument recently put forward to suggest that ultracompact configurations might be nonlinearly unstable~\cite{Keir:2014oka}~\footnote{Similar arguments have also been recently used to suggest that the superradiant instability could lead to turbulent states~\cite{Okawa:2014nda}.}.

In the following we will test some of these consequences by computing the modes of ultracompact configurations
and the time evolution of wavepackets in the vicinities of such objects.
We consider two different ultracompact objects -- constant density stars and ``gravastars'' -- briefly described below.
Our results apply also to ultracompact boson stars, which have been recently built in Ref.~\cite{Macedo:2013qea}, or to any other ultracompact object, as will become apparent from the technical details we present.
\subsection{Constant-density stars}\label{subsec:constantdensity} 
Constant-density stars are excellent idealized models to explore the properties of ultracompact objects. Because of the simplicity of the model, the metric is known analytically in the entire space. Outside the star, the spacetime is described by the Schwarzschild metric. Inside the star, the metric coefficients are given by \cite{Shapiro:1983du}
\begin{eqnarray}
f(r)&=&\frac{1}{4 R^3}\l(\sqrt{R^3-2Mr^2}-3R\sqrt{R-2M}\r)^2\,, \label{fstar}\\
B(r)&=&\l(1-\frac{2 M r^2}{R^3}\r)^{-1}\,, \label{Bstar}
\end{eqnarray}
where $R$ is the radius of the star. The pressure is given by
\begin{equation}
 p(r)=\rho_c\frac{\sqrt{3-8 \pi  R^2 \rho_c}-\sqrt{3-8 \pi  r^2 \rho_c}}{\sqrt{3-8 \pi  r^2 \rho_c}-3 \sqrt{3-8 \pi  R^2 \rho_c}}\,, \label{Pstar}
\end{equation}
where $\rho_c=3M/(4\pi R^3)$ is the density of the uniform star.

\subsection{Thin-shell gravastars}
``Gravitational condensate stars'', or gravastars, have been devised to mimic BHs~\cite{Mazur:2001fv}. In these models, the spacetime is assumed to undergo a quantum phase transition in the vicinity of the would-be BH horizon. The latter is effectively replaced by a transition layer and the BH interior by a
segment of de Sitter space \cite{Mazur:2004fk}. The effective negative pressure of the de Sitter interior contributes to sustain the self-gravity of the object for any compactness. In the static case these models have been shown to be thermodynamically~\cite{Mazur:2001fv} and dynamically~\cite{Visser:2003ge,Chirenti:2007mk,Pani:2009ss} stable for reasonable equations of state.

Here we focus on the simplest static thin-shell gravastar model, whose exterior metric for $r>R$ is identical to Schwarzschild whereas the interior, $r<R$, is described by a de Sitter metric,
\begin{equation}
f(r) = B(r)^{-1}=1-\frac{2M}{R}\frac{r^2}{R^2}\,,
\end{equation}
where $M$ is the gravastar mass measured by an observer at infinity and the effective cosmological constant of the de Sitter region is $\Lambda\equiv 6M/R^3$. The junction
conditions at $r=R$ surface have already been partially chosen by
requiring the induced metric to be continuous across the shell (cf. Ref.~\cite{Pani:2009ss} for details).
Israel's junction conditions~\cite{Israel:1966rt} then relate the discontinuities in the metric coefficients to the surface
energy $\Sigma$ and surface tension $\Theta$ of the shell as \cite{Visser:2003ge}
\begin{equation}
[[B^{-1/2}]] =-4\pi R \Sigma\,,\quad 
\left[\left[ \frac{f'B^{-1/2}}{f}\right]\right]= 8\pi (\Sigma-2\Theta)\,. \label{eq:SigmaTheta}
\end{equation}
where the symbol ``[[ ...]]'' denotes the ``jump'' in a given quantity across
the spherical shell. In the simplest model considered here, the coefficient $B$ is continuous across the shell, and therefore $\Sigma=0$, whereas the surface tension is nonzero.

\section{Perturbations of ultracompact objects}\label{model}
Various classes of perturbations of the metric (\ref{metricspherical}) are described by a master equation
\begin{equation}
\left[\frac{\partial^2}{\partial t^2}-\frac{\partial^2}{\partial r_*^2}+V_{sl}(r)\right]\Psi(r,t)=0 \,,\label{master}
\end{equation}
where $\frac{\partial^2}{\partial r_*^2}=\frac{f}{B} \frac{\partial^2}{\partial r^2}+\frac{f}{2B}(\frac{f'}{f}-\frac{B'}{B})\frac{\partial}{\partial r}$ and
\begin{eqnarray}
V_{sl}(r)&=&f\left[\frac{l(l+1)}{r^2}+\frac{1-s^2}{2r B}\left(\frac{f'}{f}-\frac{B'}{B}\right)\right.\nn\\
&&\left.+8\pi(p_{\rm rad}-\rho) \delta_{s2}\right]\,,\label{potentialmaster}
\end{eqnarray}
where the prime denotes derivative with respect to the coordinate $r$, which is related to the tortoise coordinate $r_*$ through $dr/dr_*=\sqrt{f/B}$.
In the potential~\eqref{potentialmaster} $l\geq s$, $s=0,1$ for test Klein-Gordon and Maxwell fields, respectively, whereas $s=2$ for axial perturbations of a (generically anisotropic) fluid in General Relativity (where $p_{\rm rad}=T_r^r$ and $\rho=-T_t^t$ are the radial pressure and the energy density of the fluid, respectively). In the latter case, using the field equations, the potential above reduces to
\begin{eqnarray}
 V_{2l}(r)&=&f\left[\frac{l(l+1)}{r^2}-\frac{6m(r)}{r^3}-4\pi (p_{\rm rad}-\rho)\right]\,,\label{potaxial}
\end{eqnarray}
where $m(r)$ is defined through $B(r)=(1-2m(r)/r)^{-1}$. 
Clearly, assuming a time dependence $\Psi(r,t)=\Psi(r)e^{-i\omega t}$, the radial function $\Psi$ satisfies a Schrodinger-like equation, $d^2\Psi/dr_*^2+[\omega^2-V_{sl}(r)]\Psi=0$. 

For a thin-shell gravastar, the gravitational perturbations in the interior of the star are described by the potential \eqref{potaxial}, with $-p_{\rm rad}=\rho=\Lambda/(8\pi)$ and $m(r)=M(r/R)^3$. In this case the Schroedinger-like problem in the interior simplifies considerably and can be solved analytically in terms of hypergeometric functions $F[a,b,c;z]$~\cite{Pani:2009ss}
\beq
\Psi(r)&=&
r^{l+1}(1-C(r/2M)^2)^{i\frac{M\omega}{\sqrt{C}}}\nonumber \\
&& F \bigg[\frac{l+2+i\frac{2M\omega}{\sqrt{C}}}{2},
\frac{l+1+i\frac{2M\omega}{\sqrt{C}}}{2},l+\frac{3}{2};\frac{Cr^2}{4M^2}\bigg],\quad\quad
\label{eq:interior_sol}
\eeq
where $C=(2M/R)^3$. The master function above describes \emph{both} gravitational axial and polar perturbations of the gravastar interior and has to be matched with the Regge-Wheeler or Zerilli function in the Schwarzschild exterior using suitable junction conditions~\cite{Pani:2009ss}.
\section{Long-lived modes of ultracompact objects}\label{eikonal}
%
\begin{figure}[t]
\begin{center}
\begin{tabular}{cc}
\epsfig{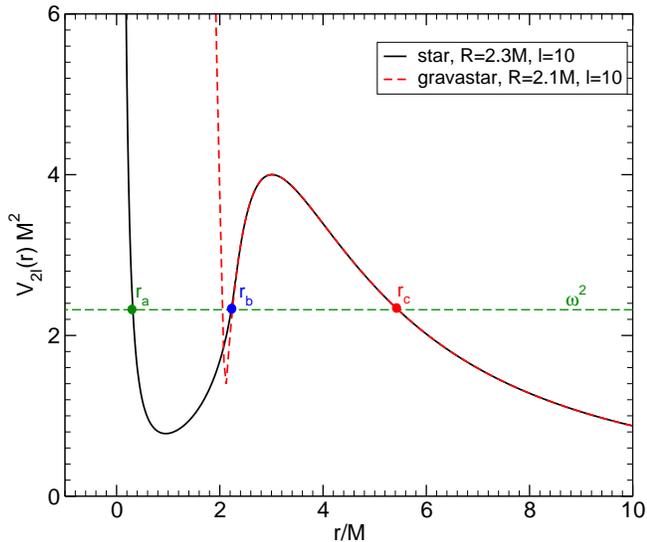}
\end{tabular}
\end{center}
\caption{\label{fig:potential}
Examples of the potential governing linear perturbations of a static ultracompact star. The black solid line and the red dashed line correspond to $l=10$ gravitational axial perturbations of a uniform star with $R=2.3 M$ and of a gravastar with $R=2.1M$, respectively.}
\end{figure}
%
\subsection{A WKB analysis}
As previously discussed, ultracompact stars have two light rings. From a point of view of massless
fields, which propagate as null particles in the eikonal regime, the light rings effectively confine
the field and give rise to long-lived modes. Before analyzing in some detail each of the specific geometries, let us perform a WKB analysis of these trapped modes.

The effective potential for wave propagation, $V_{sl}(r)$,
shares many similarities with the geodesic potential $V_{\rm geo}(r)$ to which it reduces in the eikonal limit~\cite{Cardoso:2008bp}:
it has a local maximum, diverges at the origin  and is constant at infinity.
Examples of the effective potential $V_{sl}(r)$ are shown in Fig.~\ref{fig:potential}, corresponding to $l=10$ gravitational axial perturbations of a uniform star with compactness $M/R\sim0.435$ (black solid curve) and of a thin-shell gravastar with compactness $M/R\sim 0.476$ (dashed red curve), respectively.

Because the potential necessarily develops a local minimum, it is possible to show that in the eikonal limit ($l\gg1$) the spectrum contains long-lived modes whose damping time grows exponentially with $l$. In order to do so, we follow closely the analysis by Festuccia and Liu~\cite{Festuccia:2008zx,Berti:2009wx} \footnote{These authors study the Schwarzschild-anti-de Sitter geometry, for which $V_{sl}(r)$
shares many of the properties above: it diverges at the boundaries, vanishes near the horizon and always displays a maximum at the unstable light ring.}. 
\begin{figure}[t]
\begin{center}
\begin{tabular}{cc}
\epsfig{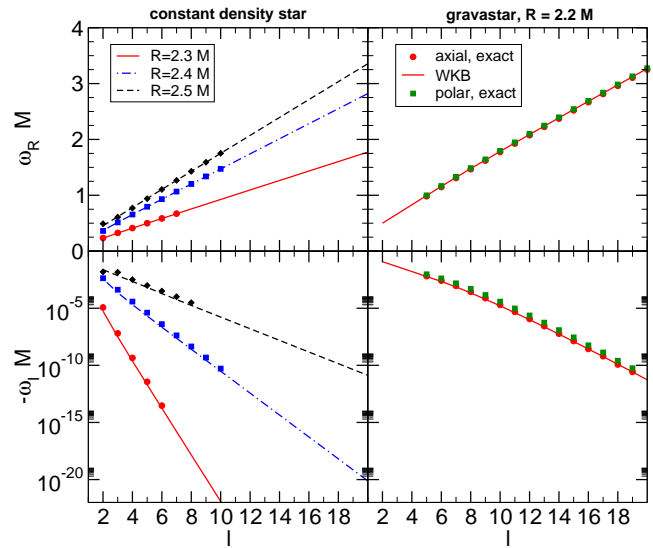}
\end{tabular}
\end{center}
\caption{\label{fig:star}
Real and imaginary parts of the long-lived modes of a uniform star for different compactness (left panels) and for a gravastar with $R=2.2M$ (right panels). Solid lines are the WKB results, whereas markers show the numerical points (when available) obtained using direct integration or continued fractions. For uniform stars we show gravitational axial modes, whereas for gravastar we show both axial modes (red circles) and gravitational polar modes with $v_s=0.1$ (green squares), where $v_s$ is related to the speed of sound on the shell~\cite{Pani:2009ss}. Note that the modes of a static gravastar become isospectral in the high-compactness regime~\cite{Pani:2009ss}.
}
\end{figure}

In the eikonal limit the potential can be approximated as $V_{sl}(r)\sim l^2f/r^2$. Let us define $r_a$, $r_b$ and $r_c$ to be the three real turning points of $\omega_R^2-V_{sl}(r)=0$ as shown in Fig.~\ref{fig:potential} for the black solid curve. When such turning points exist, the real part of the frequency of a class of long-lived modes in four spacetime dimensions is given by the WKB condition (see also Ref.~\cite{Gurvitz:1988zz})
\begin{equation}
\int_{r_a}^{r_b}\frac{dr}{\sqrt{f/B}}\sqrt{\omega_R^2-V_{sl}(r)}=\pi\left (n+1/2\right),\label{bohrsommerfeld}
\end{equation}
where $n$ is a positive integer and we have used the fact that $dr_*=dr/\sqrt{f/B}$.
The imaginary part of the frequency $\omega_I$ of these modes is given by
\be
\omega_I=-\frac{1}{8\omega_R\gamma} e^{-\Gamma}\,, \label{omI}
\ee
where
\begin{eqnarray}
\Gamma&=& 2\int_{r_b}^{r_c}\frac{dr}{\sqrt{f/B}} \sqrt{V_{sl}(r)-\omega_R^2}\,,\\
\gamma&=& \int_{r_a}^{r_b}\frac{dr}{\sqrt{f/B}} \frac{\cos^2\chi(r)}{\sqrt{\omega_R^2-V_{sl}(r)}} \,,\\
\chi(r)&=& -\frac{\pi}{4}+\int_{r_a}^{r}\frac{dr}{\sqrt{f/B}} \sqrt{\omega_R^2-V_{	sl}(r)}\,.
\end{eqnarray}
By expanding Eqs.~\eqref{bohrsommerfeld} and~\eqref{omI}, one can show that, to leading order in the eikonal limit, the mode frequency reads
\begin{equation}
\omega\sim a\, l -i\, b\, e^{-c l}\qquad l\gg1\,, \label{omega}
\end{equation}
where $a$, $b$ and $c$ are positive constants. By expanding Eq.~\eqref{bohrsommerfeld} near the minimum of the potential displayed in Fig.~\ref{fig:potential}, it is possible to show that
\begin{equation}
 a\sim\Omega_{\rm LR2}\equiv\frac{\sqrt{f(r_{\rm LR2})}}{r_{\rm LR2}}\,,
\end{equation}
where $\Omega_{\rm LR2}$ is the angular velocity of the \emph{stable} null geodesic at the light-ring location $r=r_{\rm LR2}$.
For constant-density stars this orbital frequency reads 
\begin{equation}
\Omega_{\rm LR2}=\frac{2\sqrt{M ( R-9 M/4)}}{R^2}\,,
\end{equation}
and is vanishing in the Buchdahl limit $R\to 9M/4$. For gravastars
\begin{equation}
\Omega_{\rm LR2}=\frac{\sqrt{R-2M}}{R^{3/2}},
\end{equation}
and is vanishing at the Schwarzschild limit $R\to 2M$.

\subsection{Numerical results: the spectrum of linear perturbations}

A numerical computation of the quasinormal mode (QNM) frequencies~\cite{Berti:2009kk} shows that long-lived modes are indeed part of the spectrum,
as indicated by the WKB analysis. In Fig.~\ref{fig:star} we present some of these modes for constant-density ultracompact stars with $R/M=2.3,\,2.4,\,2.5$ (left panels) and for a thin-shell gravastar with $R=2.2M$ (right panels). The exact numerical values obtained via direct integration and continued fractions (cf. e.g. Ref.~\cite{Berti:2009kk} for details) are denoted by markers and are compared against the WKB prediction (solid lines). These independent computations are in very good agreement, validating each other. 

For uniform stars (left panels of Fig.~\ref{fig:star}) we present the gravitational axial modes which are governed by the effective potential in Eq.~\eqref{potaxial}.
The existence of trapped modes in ultracompact stars was discovered in Ref.~\cite{1991RSPSA.434..449C} (see also~\cite{Andersson:1995ez,Kojima:1995nc} and \cite{Kokkotas:1999bd} for a review). Our analysis perfectly agrees with previous results and extends the latter in the case of large values of $l$.

For gravastars (right panels of Fig.~\ref{fig:star}) we present both gravitational axial and gravitational polar perturbations. The latter depends on the equation of state of the thin-shell through the parameter $v_s^2\equiv \partial\Sigma/\partial\Theta$, which is related to the speed of sound on the shell. To compute the gravastar modes we matched the exact solution~\eqref{eq:interior_sol} to the Regge-Wheeler or the Zerilli function in the Schwarzschild exterior for axial or polar modes, respectively, as discussed in detail in Ref.~\cite{Pani:2009ss}.

We note that the critical value of $l$ for which the behavior~\eqref{omega} sets in depends strongly on the compactness: the larger the star radius (constrained to $R/M\lesssim 3$) the larger the critical value of $l$. Nonetheless, the qualitative behavior is largely independent of the compactness, the nature of the modes and even the nature of the ultracompact object, as long as the latter is compact enough to support long-lived modes. In particular, our results show that trapped modes also exist in the polar sector of gravitational perturbations, which are coupled to the fluid perturbations~\cite{Kokkotas:1999bd} and that dominate the linear response of the object to external sources.

\begin{figure}[thb]
\begin{center}
\begin{tabular}{cc}
 \epsfig{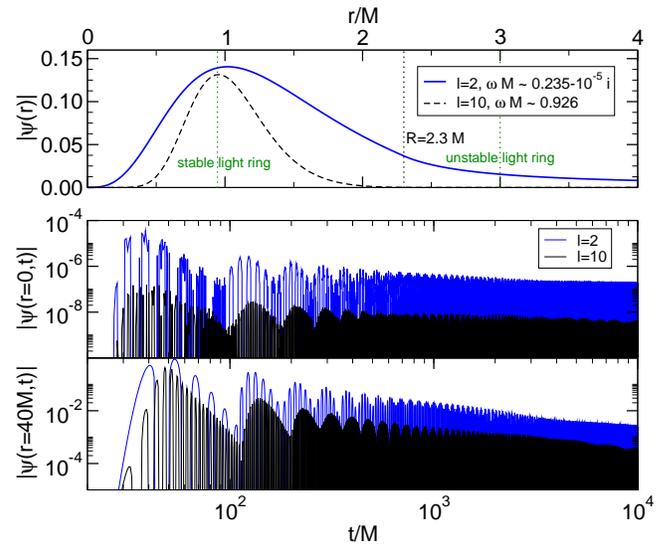}
\end{tabular}
\end{center}
\caption{\label{fig:eigenandtime}
Top panel: gravitational axial eigenfunctions of an ultracompact star for
 $l=2$ and $l=10$. The radius of the star, $R=2.3M$, is marked by a
 vertical line. High-$l$ modes correspond to eigenfunctions which
 are localized near the stable light ring. Middle and bottom panels: time evolution of
 a scalar Gaussian wavepacket with width $\sigma=4M$ centered at $r_0=6M$ in
 the background of a constant-density star of radius $R=2.3M$ for $l=2$
 and $l=10$. The waveform extracted
 at $r=0$ (middle panel) and $r=40M$ (bottom panel). Note that the Schwarzschild ringdown phase lasts until $t\sim 60 M$.}
\end{figure}

In the top panel of Fig.~\ref{fig:eigenandtime} we show a representative example of the eigenfunctions corresponding to the long-lived modes of an ultracompact object. This plot refers to a uniform star with $R=2.3M$, but different choices of the compactness and different models give similar results. The eigenfunctions are confined within the unstable light ring and within the star. Furthermore, they peak close to the location of the stable light ring and high-$l$ eigenfunctions are more and more localized around $r\sim r_{\rm LR2}$. It will be important in the following (cf. Sec.~\ref{sec:nonlinear}) to observe that the eigenfunctions spread over a distance $R/l$ in the angular direction and $\sim l^{-1/2}$ in the radial direction.

%
\subsection{Numerical results: time evolution of wavepackets}
In the middle and bottom panels of Fig.~\ref{fig:eigenandtime} we summarize the evolution of a Gaussian scalar wavepacket in the background of an ultracompact constant-density star. Initially the wavepacket is localized outside the star and has the form
\be
\dot{\Psi}(0,r)=\exp{\left[-\frac{(r + 2 \log{(r-R)}-r_0)^2}{\sigma^2}\right]}\,.
\ee
where $r_0$ and $\sigma$ denote the initial position and the width of the packet. The overdot denotes time derivative. 
\subsubsection{Imprints of the Schwarzschild BH geometry on ultracompact stars\label{QNMstraw}}
As shown in Fig.~\ref{fig:eigenandtime}, the signal initially consists of a damped sinusoid, whose frequency and damping time match closely the 
quasinormal frequencies of the {\it Schwarzschild BH} spacetime~\cite{Berti:2005ys,Berti:2009kk}.
Thus, although the QNMs of Schwarzschild BHs are {\it not} part of the spectrum
of this ultracompact star, they are still excited at early times and are an important part of the response
of this system. Such interesting ``mode camouflage'' phenomenon was observed earlier in the context of BHs surrounded by matter~\cite{Barausse:2014tra,Barausse:2014pra}.
In the present context, it also has a natural interpretation: the modes of BHs ``live'' on the external null circular geodesic~\cite{Cardoso:2008bp},
which is also present for ultracompact stars. Accordingly, we expect the BH ringdown stage to dominate until other scales become important,
in our case, after fluctuations cross the star.

This feature has two important consequences for gravitational-wave astronomy and for attempts at
proving or ruling out the existence of BHs. Any spacetime which -- close to the unstable null circular geodesic -- resembles the Kerr geometry
is expected to ringdown like a Kerr BH at early times. In other words, both dirty BHs and ultracompact stars will show
a dominant ringdown stage which is indistinguishable from that of vacuum Kerr BHs. 
This was observed for dirty BHs in Ref.~\cite{Barausse:2014tra,Barausse:2014pra} and our results show that it holds even 
for ultracompact objects, which can be looked at as a deformed BH with no horizon.
Thus, current gravitational-wave ringdown searches which {\it assume} the source is described by the Kerr geometry~\cite{Abbott:2009km,Aasi:2014bqj} are most likely
to perform well under any circumstances.

These results also have an impact on proposed methods to discriminate between BHs and other objects.
These proposals typically hinge on the no-hair theorem and the characteristic oscillation modes of these objects~\cite{Berti:2006qt}. The argument is that different objects have different oscillation modes, and the modes of BHs are known very accurately; thus, the measurement of these modes can be used to infer which object is oscillating. 
While the reasoning is correct, in practice the ringdown mode of any object which is compact enough
will be dominated at early times by a universal ringdown: it is a superposition of the QNMs of a vacuum BH.

Furthermore, it is commonly believed that different boundary conditions (for example due to the presence of an event horizon instead that of a surface) would drastically change the spectrum of ringdown modes. While it is true that the full QNM spectrum (as obtained in the frequency domain) is strongly affected by the boundary conditions, nonetheless the early-time behavior of the waveforms is mostly dominated by the macroscopic ``local'' properties of the object (i.e. by the geometry near the unstable light ring), irrespectively of the existence of a horizon~\cite{Barausse:2014tra,Barausse:2014pra}.
It is still possible --~though probably more challenging~-- to dig out the signal in the late-time stage, which will contain the object's true modes, but this would require large signal-to-noise detections~\cite{Barausse:2014tra,Barausse:2014pra}.
\subsubsection{Long-lived perturbations}

\begin{figure*}[t]
\begin{center}
\begin{tabular}{cc}
\epsfig{file=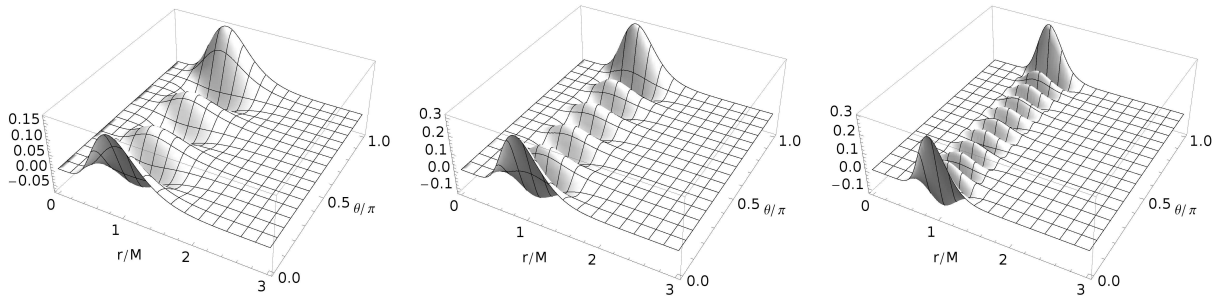,width=16cm,angle=0,clip=true}
\end{tabular}
\end{center}
\caption{\label{fig:3D}
Scalar eigenfunctions of an ultracompact star with $R=2.3M$ for $m=0$ and $l=6,10,20$ (from left to the right).
We find that the eigenfunctions have a typical width that scales as $l^{-1}$ in the angular direction and a width in the radial direction
that depends on the model used for the star, but typically ranges between $l^{-0.4}-l^{-0.8}$. Therefore, the ``aspect ratio'' of the perturbation $\sim l^{0.6}-l^{0.2}$ grows in the large-$l$ limit and the perturbation becomes more and more elongated along the radial direction.
}
\end{figure*}

The mode camouflage phase we just described lasts roughly $60M$, which corresponds to
the (roundtrip) light-crossing time for the star under consideration. The light crossing time seems to be decisive in the low-frequency modulation of the signal. At very late times, the modes of the system set in and the field decays very slowly. 
The decay rate depends on the initial conditions and on the model, but it is always slower than $1/t$. 
For example, for the case shown in the bottom panels of Fig.~\ref{fig:eigenandtime} we estimate the decay to be at most $\sim t^{-0.4}$ inside the star for the $l=10$ mode {\it assuming} it is a power-law decay. The reason why the signal decays so slowly at late-times is apparent from the top panel of Fig.~\ref{fig:eigenandtime} and also in Fig.~\ref{fig:3D}: the corresponding eigenfunctions in the frequency domain are trapped inside the star and localized near the stable light ring. 

%
\section{Spinning ultracompact objects and the ergoregion instability}
The long-lived modes that generically exist for any static ultracompact star can turn unstable when the star is spinning. This instability is related to the \emph{ergoregion instability} which affects any spacetime possessing an ergoregion but not a horizon~\cite{1978CMaPh..63..243F}. The ergoregion is defined as the spacetime region in which observers must be dragged along with rotation and cannot remain at rest. This corresponds to the timelike Killing vector $\xi_t$ becoming spacelike, i.e.
\begin{equation}
 \xi_t\cdot \xi_t=g_{tt}(r,\theta)>0\,. \label{defER}
\end{equation}
In fact, the existence of long-lived modes in the static limit is \emph{the} underlying reason of the ergoregion instability. This has been first discussed by Comins and Schutz, who studied a scalar field propagating in a slowly-rotating background in the eikonal limit~\cite{Comins}. They considered the line element
\begin{equation}
 ds^2=-F(r)dt^2+B(r) dr^2+r^2 d\theta^2+r^2 \sin^2\theta(d\phi-\varpi(r) dt)^2\,, \label{ds2rotapprox}
\end{equation}
which, although not being a solution of Einstein's equations coupled to a fluid, should approximate the exact metric describing a spinning star in the case of slow rotation and high compactness~\cite{Comins}. In such metric, the ergoregion is defined by
\begin{equation}
 \varpi(r)\sin\theta>\frac{\sqrt{F(r)}}{r}\,,
\end{equation}
and its boundary, the ergosphere, is topologically a torus.
In the eikonal limit, the Klein-Gordon equation in the background~\eqref{ds2rotapprox} can be written in the form~\cite{Comins}
\begin{equation}
 \Psi''+m^2 \frac{B}{F}(\bar{\omega}+V_+)(\bar{\omega}+V_-)\Psi=0\,, \label{KGrot}
\end{equation}
where $\bar{\omega}=\omega/m$ is a rescaled frequency, $m$ is the azimuthal number associated to the axisymmetry of the background, and
\begin{equation}
 V_\pm=-\varpi\pm \frac{\sqrt{F}}{r}\,,
\end{equation}
are the effective potentials that describe the motion of (counter-rotating for the plus sign and co-rotating for the minus sign) null geodesics in the equatorial plane of the geometry~\eqref{ds2rotapprox}.

Now, the boundary of the ergoregion (if it exists) corresponds to two real roots of $V_+=0$ and $V_+<0$ inside the ergoregion. Because $V_+\to+\infty$ at the center and attains a positive finite value in the exterior, it is clear that the ergoregion must contain a point in which $V_+$ displays a (negative) local minimum. This simple argument shows the important result that the presence of an ergoregion in a horizonless object implies the existence of \emph{stable} counter-rotating photon orbits. 

Furthermore, Eq.~\eqref{KGrot} supports unstable modes whose instability time scale in the eikonal limit grows exponentially, $\tau\equiv 1/\omega_I\sim 4\alpha e^{2\beta m}$, where $\alpha$ and $\beta$ are two positive constants~\cite{Comins}. This instability can be understood from the fact that the corresponding modes are localized near the stable photon orbit, which is situated within the ergosphere, and are confined within the star. This confinement provides the arena for the instability to grow through the negative-energy states that are allowed within the ergoregion~\cite{1978CMaPh..63..243F}. Likewise, this argument also explains why spinning BHs --~that also possess a light ring and an ergoregion~-- are linearly stable, because the presence of the horizon forbids the existence of trapped modes.

Although the analysis of Ref.~\cite{Comins} is approximate, such result has been subsequently extended to low values of $(l,m)$~\cite{YoshidaEriguchi} and to gravitational axial perturbations~\cite{Kokkotas:2002sf}. In both cases, the instability time scale has been found to be much shorter, ranging from seconds to minutes for low-$m$ gravitational perturbations of uniform constant stars~\cite{Kokkotas:2002sf}. The conclusion of these studies is that, if long-lived modes exist in the static case, they become unstable for sufficiently high rotation rates. The onset of the instability precisely corresponds to the appearance of an ergoregion in the interior of an ultracompact star~\cite{Kokkotas:2002sf}. The same picture applies to other ultracompact objects such as gravastars and boson stars, which become linearly unstable when they possess an ergoregion~\cite{Cardoso:2007az} with an instability time scale that depends strongly on the compactness~\cite{Chirenti:2008pf}. The same instability affects also Kerr-like BH geometries spinning above the Kerr bound (so-called superspinars~\cite{Cardoso:2008kj}) when the dissipation at the horizon is not enough to quench the negative-energy states trapped within the ergoregion~\cite{Pani:2010jz}. Finally, the ergoregion instability of acoustic geometries was recently reported~\cite{Oliveira:2014oja,Hod:2014hda}.

\section{The nonlinear regime}\label{sec:nonlinear}
The argument for nonlinear instability given earlier is anchored on the large
lifetimes of linear fluctuations. This argument carries over equally to other more familiar
contexts, e.g. to conservative systems with normal modes. Generically however, normal-mode systems
are an idealization and neglect {\it any} form of dissipation.
The outstanding feature of ultracompact stars is that gravitational-wave dissipation is already included and is negligible.

We can foresee at least two possible outcomes for the nonlinear development of ultracompact stars;
which one is actually chosen depends on the details of the object's composition:

\noindent {\bf I. Other dissipation mechanisms are relevant}, in which case the star is stable. 
Loss of energy through gravitational-wave emission is suppressed for ultracompact stars,
but this is not the only dissipation mechanism. For example, viscosity in neutron stars plays an important role on relatively short time scales,
and may quench possible nonlinear instabilities for very compact stars. Simple expressions for the dissipative time scales as functions of the angular number $l$ and the parameters of a neutron star were derived in Ref.~\cite{1987ApJ...314..234C}:
\begin{eqnarray}
\tau_\eta&=&\frac{10}{(l-1)(2l+1)} \rho_{14}^{-5/4} T_5^{2} \left(\frac{R}{4.5\,{\rm km}}\right)^{2} \,{\rm s}\,,\\
\tau_\kappa&=& 10^{14}\tau_\eta \frac{(l-1)^2}{l^3} \rho_{14}^{19/12} T_5^{-2} \left(\frac{R}{4.5\,{\rm km}}\right)^{2}\,,\\
\tau_\zeta&>&61 \tau_\eta \frac{\eta}{\zeta}\,,
\end{eqnarray}
where $\rho_{14}=\rho/(10^{14}\,{\rm g/cm}^3)$, $T_5=T/(10^5 K)$, $T$ is the neutron-star temperature, $\tau_\eta$, $\tau_\kappa$ and $\tau_\zeta$ are the time scales for shear viscosity, thermal conductivity and bulk viscosity, respectively, whereas $\eta$, $\zeta$ and $\kappa$ are dissipation coefficients. 

These are order-of-magnitude estimates, valid in principle only for neutron stars. Any hypothetical ultracompact
star will however also be affected by dissipation of this nature, whose time scale becomes shorter at shorter scales, i.e., larger $l$. Note, however, that some modes are only weakly coupled with the fluid perturbations (e.g. gravitational axial modes and $w$-modes in general~\cite{Kokkotas:1999bd}) so that only a small fraction of the energy contained in such modes can be dissipated through viscosity. Furthermore, the interior of exotic ultracompact stars could be made of a superfluid as in self-gravitating Bose-Einstein condensates~\cite{Liebling:2012fv} and also in this case viscosity is expected to be negligible.

\noindent{\bf II. Nonlinear effects become relevant.} Let us now assume that there is no dissipation mechanism strong enough
to damp linear perturbations on realistic time scales. Recent studies of gravitational collapse of small scalar-field wavepackets in anti-de Sitter geometries (which are another example of conservative systems with normal modes at linear level), suggest that broad classes of initial data
{\it always} collapse to form BHs, through a ``weakly turbulent'' mechanism~\cite{Bizon:2011gg}. The process, still not well understood,
involves blueshift of initial perturbations which eventually collapse to a small BH. If active for ultracompact stars,
it most likely involves a growth of curvature close to the stable null geodesic and consequent collapse to small BHs. The number of such small BHs would be tied to the angular number of the mode in question and would scale as $l$. For large enough initial fluctuation, the BHs that would form can be large enough to swallow the star in less than a Hubble time.

\begin{figure}[th]
\begin{center}
\begin{tabular}{cc}
\epsfig{file=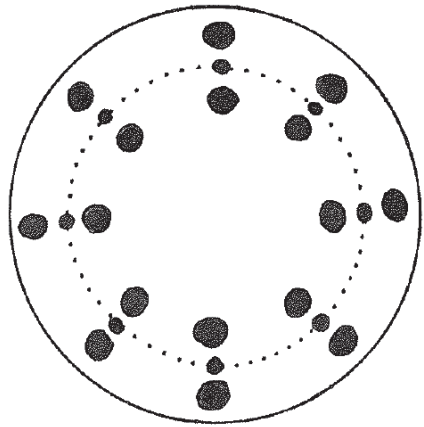,width=6cm,angle=0,clip=true}
\end{tabular}
\end{center}
\caption{\label{fig:fragmentation}
Pictorial description of the nonlinear evolution of a perturbed ultracompact object. 
The figure represents the equatorial density profile of the object.
The solid circumference represents the unperturbed surface, whereas the dashed line represents the stable light ring at its interior.
Solid circles represent condensation of nonlinear-growth structures which are the bi-product of the DCF instability.
The core is left unperturbed and is now a less compact --~and therefore stable~-- configuration. Likewise, the
the solid circles are also stable and subsequent time evolution presumably leads to a fall-back on the core.
Gravitational radiation, generated during this and subsequent repetitions of this process will lead to
loss of mass and possibly a reduction of the star's compactness.
}
\end{figure}

Do nonlinear effects always conspire to produce catastrophic results?
The answer is no. Recent studies show that there exist initial data which are nonlinearly stable against such weakly turbulent mechanism~\cite{Buchel:2013uba,Dias:2012tq,Balasubramanian:2014cja}. How generic such initial conditions are is unclear at the present time. Nevertheless, a plethora of other nonlinear effects might play a role and one in particular is likely to be dominant: fragmentation via a
``Dyson-Chandrasekhar-Fermi'' (DCF) mechanism, which is akin to the Rayleigh-Plateau fragmentation of fluid cylinders~\cite{Dyson:1893:1,Dyson:1893:2,1953ApJ...118..116C,Cardoso:2006sj}. To show this point we observe that, at linear level, the eigenfunctions have a width $\sim l^{-1}$ in the angular direction $\theta$ and a width $\sim l^{-\chi}$ in the radial direction, where $\chi<1$ depends on the star model (cf. Figs.~\ref{fig:eigenandtime} and \ref{fig:3D} for a representative example of a constant-density star).
Therefore the perturbations are asymmetric, elongated along the radial direction and their elongation grows with $l$.

Let us now assume for simplicity that we are dealing with axisymmetric modes.
Axisymmetric distributions of matter such as these elongated, long-lived modes are unstable against the same DCF 
mechanism that affects thin cylinders or rings of matter~\cite{Dyson:1893:1,Dyson:1893:2,1953ApJ...118..116C,Cardoso:2006sj}. 
The minimum growth time scale of this instability scales as $\tau_{\rm DCF} \sim \delta\rho^{-1/2}$, where $\delta\rho$ is the density fluctuation.
The requirement that nonlinearities take over is that $\tau_{\rm DCF}$ be much smaller than the lifetime of linear fluctuations. Because the latter grows exponentially with $l$ for an ultracompact object, it is easy to show that fragmentation becomes important already at moderately small values of $l$ even for $\delta\rho/\rho \sim 10^{-16}$ or smaller. In other words, we are arguing that even though ``weak turbulence'' may be negligible, fragmentation instabilities are not.

The fragmentation of the linear eigenfunction leads to a configuration which can look like that depicted in Fig.~\ref{fig:fragmentation} (see also nonlinear results for fragmentation of black strings~\cite{Lehner:2010pn}): it consists on a spherically symmetric core surrounded by droplets of the star fluid, whose sizes are much smaller than that of the original star. It is easy to see that these smaller droplets, although of the same material as the original star, are much less compact because they are much smaller and are therefore expected to be themselves stable.
Likewise, the core of the star is also less compact and stable. On longer time scales, these droplets re-arrange and fall into the core, and the process continues. The dynamical picture looks like that of a ``boiling'' fluid, and radiates a non-negligible amount of radiation.
If this scenario is correct, a sizable fraction of the object's initial mass can be dispersed to infinity, possibly
reducing the compactness of the final object to values which no longer allow for the existence of light rings.

\section{Conclusions}
Strong and growing evidence suggests that supermassive compact objects in our Universe 
are BHs. Nevertheless, incontrovertible proofs are hard to come by and would likely require detection of Hawking radiation from the event horizon, the latter being negligible for astrophysical objects.
As such, fundamental mechanisms that forbid the existence of ultracompact stars are mostly welcome and would automatically
imply that (the much more easily achievable) observations of a light ring are detections of BHs in fact. There are at least two known mechanisms that might do just that.
One such mechanism is the possible nonlinear instability of any ultracompact star, which have one unstable light ring in its exterior
(and another stable light ring in its interior). We have provided additional evidence that such objects have long-lived fluctuations which may fragment the star
and make it less compact on long time scales. Alternatively, weak turbulence might lead to collapse of the star into a BH. Whether or not the instability is actually relevant depends on possible additional dissipation mechanisms.

When rotation is added, long-lived fluctuations become unstable already at the {\it linear} level. This is also known as ergoregion instability, and has been used to exclude highly spinning, compact objects~\cite{Cardoso:2007az,Cardoso:2008kj,Pani:2010jz}.
Taken together, these results suggest that the observation of the light ring alone --~a challenging task which is nevertheless
within the reach of next facilities such as, for instance, the Event Horizon Telescope~\cite{Lu:2014zja}~-- is evidence enough for the existence of BHs, a truly remarkable consequence.

Clearly, future work should consider the difficult but fundamental problem of following long-lived fluctuations through the nonlinear regime, to understand the role of dissipation, the time scale associated with possible nonlinear instabilities, and the issue of the final state.
\begin{acknowledgments}
VC and PP acknowledge the kind hospitality of the Yukawa Institute for Theoretical Physics, where part of this work has been done during the
workshop ``Holographic vistas on Gravity and Strings'', YITP-T-14-1.
The authors acknowledge financial support provided under the European Union's FP7 ERC Starting Grant ``The dynamics of black holes:
testing the limits of Einstein's theory'' grant agreement no. DyBHo--256667.
The authors would like also to thank Conselho Nacional de Desenvolvimento Cient\'ifico e Tecnol\'ogico (CNPq) and Coordena\c{c}\~ao de Aperfei\c{c}oamento de Pessoal de N\'ivel Superior (CAPES), from Brazil, for partial financial support.
This research was supported in part by Perimeter Institute for Theoretical Physics. 
Research at Perimeter Institute is supported by the Government of Canada through 
Industry Canada and by the Province of Ontario through the Ministry of Economic Development 
$\&$ Innovation.
This work was supported by the NRHEP 295189 FP7-PEOPLE-2011-IRSES Grant and by FCT-Portugal through projects IF/00293/2013 and CERN/FP/123593/2011.
Computations were performed on the ``Baltasar Sete-Sois'' cluster at IST.
\end{acknowledgments}
%
\bibliographystyle{h-physrev4}
\bibliography{Eikonal}

\end{document}